\documentclass[sigconf, nonacm]{acmart}

\usepackage{algpseudocode}
\usepackage{algorithm}
\usepackage{xspace}
\usepackage{subcaption}
\usepackage{enumitem}
\usepackage{graphicx} 
\usepackage{listings}
\usepackage{balance}

\usepackage{xcolor}
\usepackage{xstring}

\definecolor{dkgreen}{rgb}{0,0.6,0}
\definecolor{ltgray}{rgb}{0.5,0.5,0.5}

\makeatletter
\newif\ifcolname
\colnamefalse

\def\keywordcheck{%
\IfStrEq*{\the\lst@token}{select}{\global\colnametrue}{}%
\IfStrEq*{\the\lst@token}{where}{\global\colnametrue}{}%
\IfStrEq*{\the\lst@token}{from}{\global\colnamefalse}{}%
\IfStrEq*{\the\lst@token}{SEMANTIC}{\global\colnamefalse}{}%
\color{blue}%
}
\def\setidcolor{%
\ifcolname\color{purple}\else\color{black}\fi%
}
\makeatother

\usepackage{fontawesome5}
\usepackage{hyperref}

\begin{document}
\title{SSQL - Semantic SQL}
\subtitle{Combining and optimizing semantic predicates in SQL}

\author{Akash Mittal}
\affiliation{%
}
\email{akashm3@illinois.edu}

\author{Anshul Bheemreddy}
\affiliation{%
}
\email{anshulb3@illinois.edu}

\author{Huili Tao}
\affiliation{%
}
\email{huilit2@illinois.edu}

\begin{abstract}
    In recent years, the surge in unstructured data analysis, facilitated by advancements in Machine Learning (ML), has prompted diverse approaches for handling images, text documents, and videos. Analysts, leveraging ML models, can extract meaningful information from unstructured data and store it in relational databases, allowing the execution of SQL queries for further analysis. Simultaneously, vector databases have emerged, embedding unstructured data for efficient top-k queries based on textual queries. This paper introduces a novel framework \textit{SSQL - Semantic SQL} that utilizes these two approaches, enabling the incorporation of semantic queries within SQL statements. Our approach extends SQL queries with dedicated keywords for specifying semantic queries alongside predicates related to ML model results and metadata. Our experimental results show that using just semantic queries fails catastrophically to answer count and spatial queries in more than 60\% of the cases. Our proposed method jointly optimizes the queries containing both semantic predicates and predicates on structured tables, such as those generated by ML models or other metadata. Further, to improve the query results, we incorporated human-in-the-loop feedback to determine the optimal similarity score threshold for returning results. We have open-sourced our system at \href{https://github.com/akash17mittal/semantic-sql}{\faGithubSquare} 
\end{abstract}

\maketitle

\section{Introduction}

\paragraph{Background} In recent years, there has been a lot of work related to analyzing unstructured data like images, text documents, and videos. This has been possible because of the advancements in Machine Learning. For e.g., an analyst who wants to analyze traffic camera video can run an object detection model on the frames of the video and store these results in the relational database to execute SQL queries on top of this database. Another direction of work is also prevalent where unstructured data is embedded using some machine learning model and stored in the vector database. These embedded vectors hopefully capture the semantic essence of the data so that users can then do top-k queries using text that is also embedded in the same space which allows users to search for semantic information that we might not have an ML model for. This paper presents a framework that takes the best of these two approaches and supports semantic queries in the SQL statements. To support queries where the user is interested in fetching all the results, we use a user feedback loop to determine what the optimal similarity score threshold should be where we return a result. 

\paragraph{Problem} Analyzing unstructured data using ML models by storing the results in a relational database has limited capabilities when we don't have a trained ML model for a specific user query. For e.g., a user wants to retrieve images that contain both pedestrians and cars close together in snowy weather in the last 7 days. We may have an object detection model that can detect pedestrians and cars bounding boxes but we may not have a weather detection model. So, relying on just ML models is not enough. We cannot rely only on vector stores either because embeddings often lose some details such as the precise distance between objects and the exact count of objects in an image since they are typically encoded in a lower dimensional space. Therefore, they are not great for reasoning about exact numerical answers like finding the exact location of an object in an image or the exact time the image was taken. This is problematic in areas such as spatial locality e.g. finding images where pedestrians and cars are close together, or filtering based on metadata such as timestamps, etc. The proposed technique bridges this gap. 

\paragraph{Proposed Approach} We extend the standard SQL with new keywords that can be used to specify semantic queries along with the predicates related to the ML model results' tables and metadata. We propose joint optimization of queries containing both semantic predicates and predicates on structured tables like object detection or other metadata.

\paragraph{Limitations} 

The main challenge is supporting exact queries and providing guarantees. Assuming ML models give correct predictions, exact queries can be supported on ML models results table but in the case of semantic predicates, providing guarantees on the results is hard. Also, evaluating the accuracy of semantic queries is non-trivial since we lack access to ground truth data. For example, if we want to make a query to find all the images that are snowing, we need ground truth data on whether it is actually snowing in each image or not which we don't have.

\paragraph{Difference from Existing Work} 
Significant work has been done on semantic search through embedding both the query and the content in the same vector space such as the Embedding-based Retrieval method from \cite{huang2020embedding}. The approach has been shown to work even with cross-modal retrieval (such as when the query is text and the contents are images) with Google's FashionBERT \cite{gao2020fashionbert} and OpenAI's CLIP. However, to our knowledge, there isn't any significant work that combines this semantic search of unstructured data with a more traditional search of structured data by extending the SQL syntax.

\paragraph{Contributions} We have three main contributions:
\begin{enumerate}[leftmargin=*]
    \item We proposed a novel approach to combine semantic predicates in the SQL queries while ensuring the accuracy of the results using human-in-the-loop feedback.
    \item We did extensive experimentation to show the limitations of the previous research works and the effectiveness of our system.
    \item We have open-sourced our system for the research community for future development.
\end{enumerate}

\section{Problem Definition}

From our literature survey, we identified the following gaps in the analysis of unstructured data:

\paragraph{Limitations of ML models} The first problem is its dependency on trained data. ML models depend on training data to learn patterns and make predictions. If there is no trained model for a specific user query, the system cannot effectively process or understand the query, making it difficult to extract the desired insights or information from the unstructured data. For e.g., if a user is interested in fetching red car images, we may not have an ML model that predicts red cars in the images.

The second problem is ML model has limitations in dealing with multiple types of information. Some user queries require integrating information from various sources or domains. For example, the user query related to fetching "cars in snowy weather" may require running multiple ML models like object detection and weather detection model which is computationally heavy. Integrating data from diverse sources is challenging for ML models, which are typically designed for single-domain tasks. Handling multi-modal data and making connections between different data types can be complex and may require large labeled datasets.

\paragraph{Limitation of just semantic queries} To support semantic queries, we need to embed the data in the lower dimensional space. These semantic embeddings encode the high-level information present in the unstructured data like images, text, etc. Embedding the data into the lower dimensional space loses information such as precise distance between objects or count of the objects in the image. In a search task where user needs to find instances where "pedestrians and cars are separated by 10 pixels" or the query where you want to fetch records "containing more than 5 cars" semantic search alone will not be able to directly support this.
Also, it is challenging to include other metadata information like timestamps, etc. in the semantic search.

\section{Methodology}

\begin{figure*}
  \includegraphics[height=6cm]{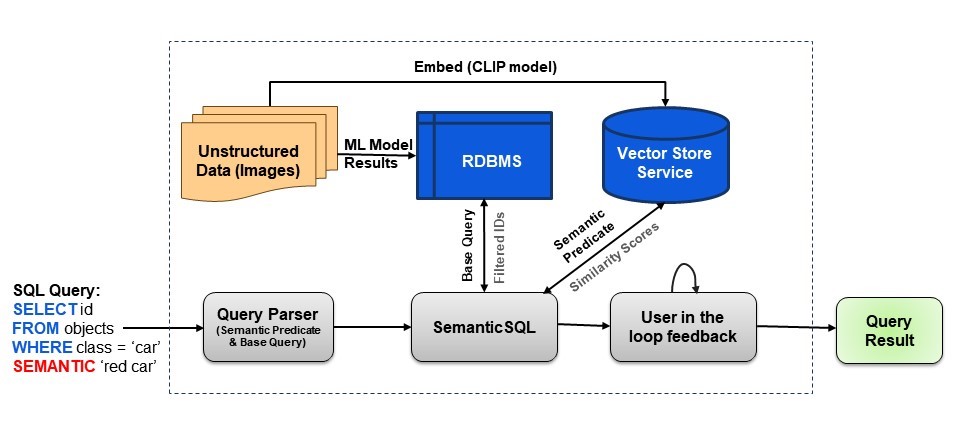}
  \caption{Semantic SQL architecture diagram}
  \label{fig:architecture}
\end{figure*}

To solve the above problems, we propose a technique that combines the best of both worlds i.e. using ML models and storing the results in RDBMS to execute SQL queries \& using vector stores for supporting queries when we don't have a ML model. We show the overall architecture of our system in the figure \ref{fig:architecture}. Our system broadly contains the following four components:
\paragraph{1. Relation Database} We utilized relational databases to store the metadata of unstructured data such as timestamp, image sizes, etc. We also store the results of the execution of the ML models in the database tables to support SQL queries. We used SQLITE database in this paper but any RDBMS can be used. An example schema of the \textit{object detection results} table is shown in the table \ref{table:1}. 

\begin{table}
\centering
\begin{tabular}{||c c||} 
 \hline
 Column Name & Column Type\\ [0.5ex] 
 \hline\hline
  image\_id & INTEGER \\
  object\_id & INTEGER  \\
  object\_class & STRING \\
  bbox\_xmin & FLOAT \\
  bbox\_xmax & FLOAT \\
  bbox\_ymin & FLOAT \\
  bbox\_ymax & FLOAT \\
 \hline
\end{tabular}
\caption{Object detection results table schema}
\label{table:1}
\end{table}

\paragraph{2. SQL Extension} We extended the standard SQL to allow users to specify semantic predicates along with the other predicates for e.g. 
\begin{lstlisting}
SELECT DISTINCT frame
FROM object_detection_results
WHERE class = 'car' AND x_max < 500 
|\color{red}SEMANTIC| = 'big green car';
\end{lstlisting}
Users can optionally specify the semantic predicate at the end of the SQL query using the keyword \textbf{SEMANTIC}. Currently, we only support one semantic predicate. We added the support of this keyword in the open sourced SQL parser named \href{https://github.com/tobymao/sqlglot}{SQLglot}.

\paragraph{3. Semantic Search/Vector Store} In order to perform cross-modal (text-to-image) semantic searches, we used OpenAI's CLIP model \cite{radford2021learning} to embed both the images and the text query into the same embedding space. We chose to use Open AI's CLIP (Contrastive Language–Image Pretraining) model for this task because of its ability to create semantically rich embeddings of both images and text. The model was also trained on a vast and diverse dataset of millions of images which allows it to create semantically rich embeddings even for images and text that it has not been explicitly trained on. Then, in order to find the image(s) that most closely match the semantic meaning of our text query, we just have to find the embedded image vectors that have the highest similarity to our embedded text query vector. There are multiple methods for computing the similarity between vectors; however, we used normalized Euclidean or L2 distance to compute these similarities. After computing this distance between all the image embeddings and the embedded text query, we can simply return the images with the lowest distance to the embedded text query. Normalization is achieved by dividing each vector by its magnitude, following the formula $\mathbf{v}^{\prime}=\frac{\mathbf{v}}{\|\mathbf{v}\|}$. Once normalized, the comparison between the text vector and each image vector is conducted using the Euclidean distance, calculated as $d(\mathbf{a}, \mathbf{b})=\sqrt{\sum_{i=1}^n\left(a_i-b_i\right)^2}$, where $\mathbf{a}$ and $\mathbf{b}$ are vectors in the $\mathrm{n}$-dimensional space. Further to reduce computation costs, we integrated vector store FAISS (Facebook AI Similarity Search) \cite{johnson2019billion} to store the image embeddings and efficiently compare the embedded text query to those embeddings.

\paragraph{4. User Feedback loop} Most of the existing vector stores support top-k queries where given a user text query, it is embedded into the lower dimensional space. Then, it is matched with the embeddings in the vector store to retrieve top-k results. But for the queries, when the user is interested in fetching all the results, we need to find the similarity threshold. The records with similarity above this threshold are returned. We use \textit{human-in-the-loop} approach to decide the threshold. We designed a user interface where we show the fetched records to users one by one at different thresholds and the user gives feedback by answering "Yes/No" if the fetched record satisfies predicates in the SQL query. We choose these samples strategically to find the optimal threshold in the minimum number of steps as shown in the algorithm \ref{alg:cap}.

\begin{algorithm}
\caption{Query Execution Algorithm}\label{alg:cap}
\begin{algorithmic}
\Require $sql\_query$
\State $parsed\_query \gets parse\_query(sql\_query)$
\If{$parsed\_query$ has SEMANTIC}
    \State $base\_sql\_query \gets remove\_semantic(parsed\_query)$
    \State $semantic\_predicate \gets extract\_semantic(parsed\_query)$
    \State $sql\_results \gets execute\_sql\_query(base\_sql\_query)$
    \State $sim\_scores \gets similarity(sql\_results, semantic\_predicate)$

    \State $results \gets \emptyset$
    
    \While{$size(sim\_scores) \geq 1$}
        \State $img \gets get\_image\_at\_50th\_percentile(sim\_scores) $
        \State $user\_feedback \gets get\_user\_feedback(img)$
        \If{$user\_feedback$ is positive}
            \State $positive\_imgs \gets imgs\_above\_50th\_per(sim\_scores)$
            \State $results \gets union(results, positive\_imgs) $
            \State $sim\_scores \gets imgs\_below\_50th\_per(sim\_scores)$
        \Else
            \State $sim\_scores \gets imgs\_above\_50th\_per(sim\_scores)$
        \EndIf
    \EndWhile
\Else
    \State $results \gets execute\_sql\_query(sql\_query)$
\EndIf

\end{algorithmic}
\end{algorithm}

 \section{Evaluation}

 Our evaluation aims to answer the following questions:
 \begin{enumerate}[leftmargin=*]
     \item How effective are the semantic queries in capturing spatial and count information?
     \item Whether combining SQL queries with semantic predicates ensures the correctness of the results?
     \item How is the user experience in providing feedback for semantic SQL queries?
 \end{enumerate}
 
 \subsection{Performance Metrics}

 \begin{enumerate}[leftmargin=*]
     \item \textbf{Accuracy:} Evaluate whether the combined queries lead to correct predictions. We will then compare the results to a ground truth or manually verified datasets. We will be assuming that the ML model used to generate object detections is 100 percent accurate by using the ground truth bounding box data as a substitution for the ML model since the exact ML model is not the focus of this paper.
     \item \textbf{User Feedback:} Collect user feedback on the system's ability to handle combined queries. Ask users about their satisfaction with the predictions and whether they are relevant to their queries.
 \end{enumerate}
 
\subsection{Dataset Details}
We evaluated our technique on the COCO dataset \cite{lin2015microsoft}. This dataset contains 330k images with 80 object categories. We randomly sampled 20k images. These sampled images contain ~145k object instances in total. Each image contains ~7 objects on average that are spatially located in the real-world context.

\subsection{Query Types}
We did both qualitative and quantitative analysis of our approach on four different types of queries and showed the results where just semantic queries fail and just SQL queries will fail.

\begin{itemize}[leftmargin=*]
    \item \textbf{Multiple objects in an image:} When assessing the performance of semantic queries for detecting multiple objects in an image, we observed limitations in capturing all relevant information. Our evaluation focused on specific semantic queries such as "${n.}$" and "${n.}$," where $n.$ represents object names such as "cat," "bottle," and "person." We analyzed the top three images with the highest matching scores and compared them with ground truth results obtained through a deterministic SQL query. Unlike semantic queries, the SQL query guarantees 100\% accuracy in object detection results. We evaluated this query for 2124 object pairs and found that only 608 pairs were able to satisfy semantic query successfully as shown in figure \ref{fig:andquery}. An example SQL query to test multiple objects is shown below.
\begin{lstlisting}
SELECT DISTINCT id from objects WHERE class_name='person' 
INTERSECT 
SELECT DISTINCT id from objects WHERE class_name= 'apple'
\end{lstlisting}

\begin{figure}
\centering
\includegraphics[width=0.5\columnwidth]{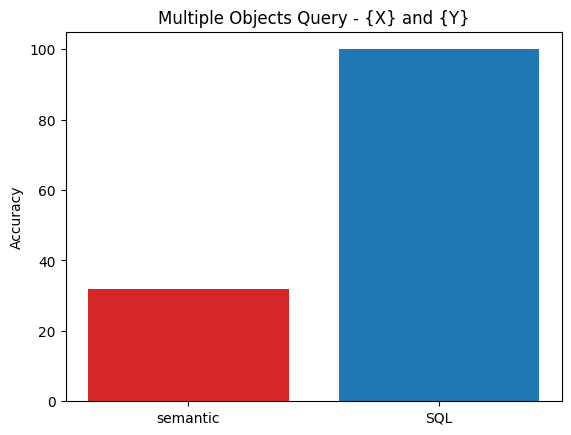}
\caption{Performance comparison on multiple objects query}
\label{fig:andquery}
\end{figure}

    \item \textbf{Count Queries:} We show the limitation of semantic queries for count queries. We evaluated it on the "${count}$ ${object}$" semantic query where $count$ can take values $one, two,..., ten$ and $object$ can take any object class. We considered the top-3 highest matching images and compare it with the ground truth result found by running SQL query. SQL query for count queries will return 100\% accurate results. We compare the performance for 2 different objects i.e., horses and cars for ten different values of count. The results are shown in the figure \ref{fig:countquery}. An example count SQL query for fetching images containing four horses is shown below:
\begin{lstlisting}
SELECT id, COUNT(*) as c
FROM objects
WHERE class_name='horse'
GROUP BY id
HAVING c = 4
\end{lstlisting}

\begin{figure}
\centering
\begin{subfigure}{.5\columnwidth}
\centering
\includegraphics[width=1.0\columnwidth]{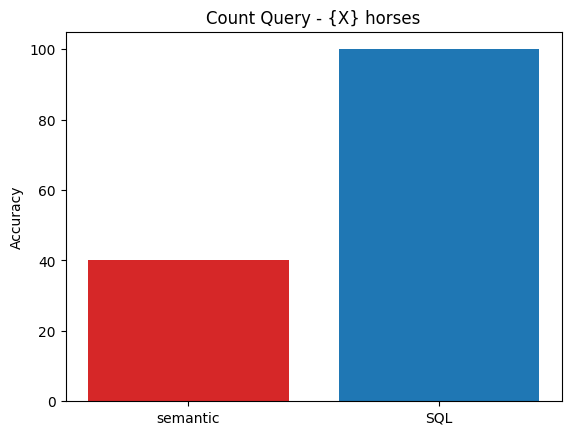}
\caption{one/two/.../ten horse/s}
\end{subfigure}%
\begin{subfigure}{.5\columnwidth}
\centering
\includegraphics[width=1.0\columnwidth]{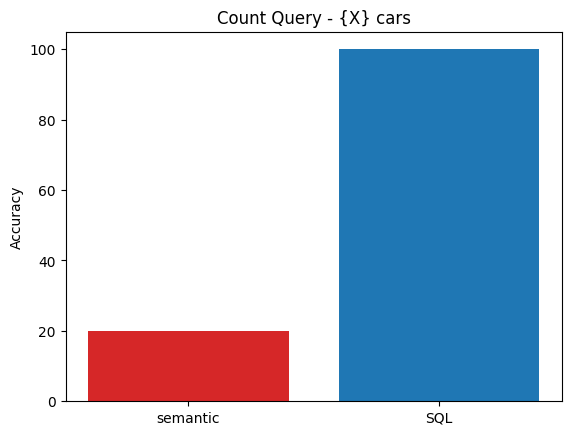}
\caption{one/two/.../ten car/s}
\end{subfigure}%
\caption{Performance comparison on count queries}
\label{fig:countquery}
\end{figure}
   
    \item \textbf{Spatial locality of objects:} When examining the spatial locality of objects, the SQL query utilizing the ground truth bounding boxes greatly outperforms the semantic queries. Our evaluation focused on a specific semantic query: "${object}$ in the top left/bottom right corner". We analyzed the top three images with the highest matching scores and compared them with the ground truth results obtained through a deterministic SQL query. The SQL query, incorporating both x and y-axis ranges, ensures 100\% accuracy in the results, as it precisely captures the spatial information of the queried objects. A SQL query to get the images containing the car in the bottom right corner is shown below. 
\begin{lstlisting}
SELECT DISTINCT id
FROM objects
WHERE class_name='car' AND x1>340 AND y1 > 340
\end{lstlisting}

\begin{figure}
\centering
\begin{subfigure}{.5\columnwidth}
\centering
\includegraphics[width=1.0\columnwidth]{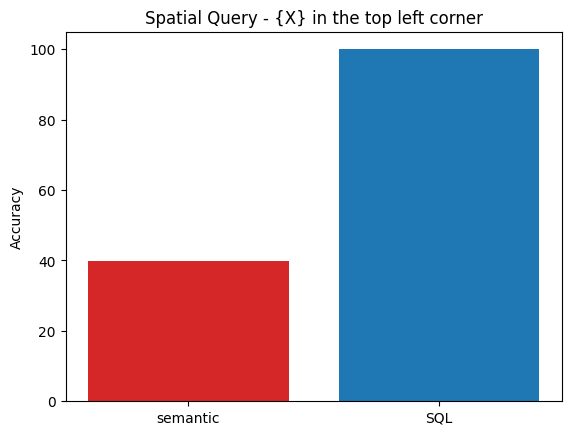}
\caption{X takes 78 different values}
\end{subfigure}%
\begin{subfigure}{.5\columnwidth}
\centering
\includegraphics[width=1.0\columnwidth]{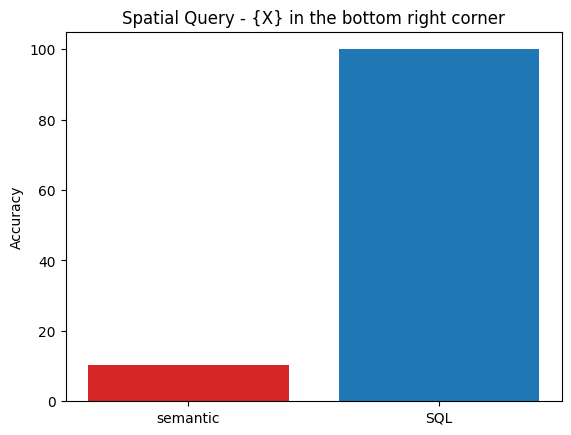}
\caption{X takes 67 different values}
\end{subfigure}%
\caption{Performance comparison on spatial locality query}
\label{fig:spatialquery}
\end{figure}

    \item \textbf {Contextual Queries:} Just SQL queries struggle with capturing the contextual relationships needed for such queries as gender, color attribute, and so on. Semantic queries such as \textit{men in suits}, with their focus on meaning, can better interpret and retrieve relevant information. We did qualitative evaluation of various semantic queries and found satisfying results. We cannot evaluate such semantic queries using just SQL because of the lack of ML models.

\end{itemize}
\subsection{Complex Semantic SQL Query Analysis}
\begin{figure}
\centering
\includegraphics[height=4cm]{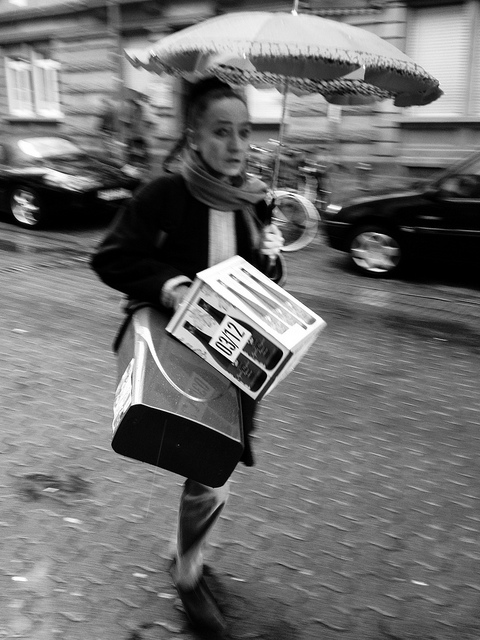}
\caption{Result returned by the SSQL}
\label{fig:complexssql}
\end{figure}

We have evaluated our system on multiple queries containing both semantic predicates and predicates on structured tables. More results can be found in the github repository. In this section, we show an example of the complex query where our proposed approach shines as compared to using just SQL and semantic query. User is interested in the images that contain \textbf{one women with umbrella and exact two cars}. The SSQL query is shown below. The results of this query are shown in the figure \ref{fig:complexssql}

\begin{lstlisting}
SELECT id
FROM
  (SELECT id, COUNT(*) AS c
   FROM objects
   WHERE class_name='person'
   GROUP BY id
   HAVING c = 1) INTERSECT
SELECT DISTINCT id
FROM objects
WHERE class_name='umbrella' INTERSECT
  SELECT id
  FROM
    (SELECT id, COUNT(*) AS c
     FROM objects
     WHERE class_name='car'
     GROUP BY id
     HAVING c = 2)
|\color{red}SEMANTIC| 'women no kids'
\end{lstlisting}

\begin{figure}
\centering
\includegraphics[width=1.0\columnwidth]{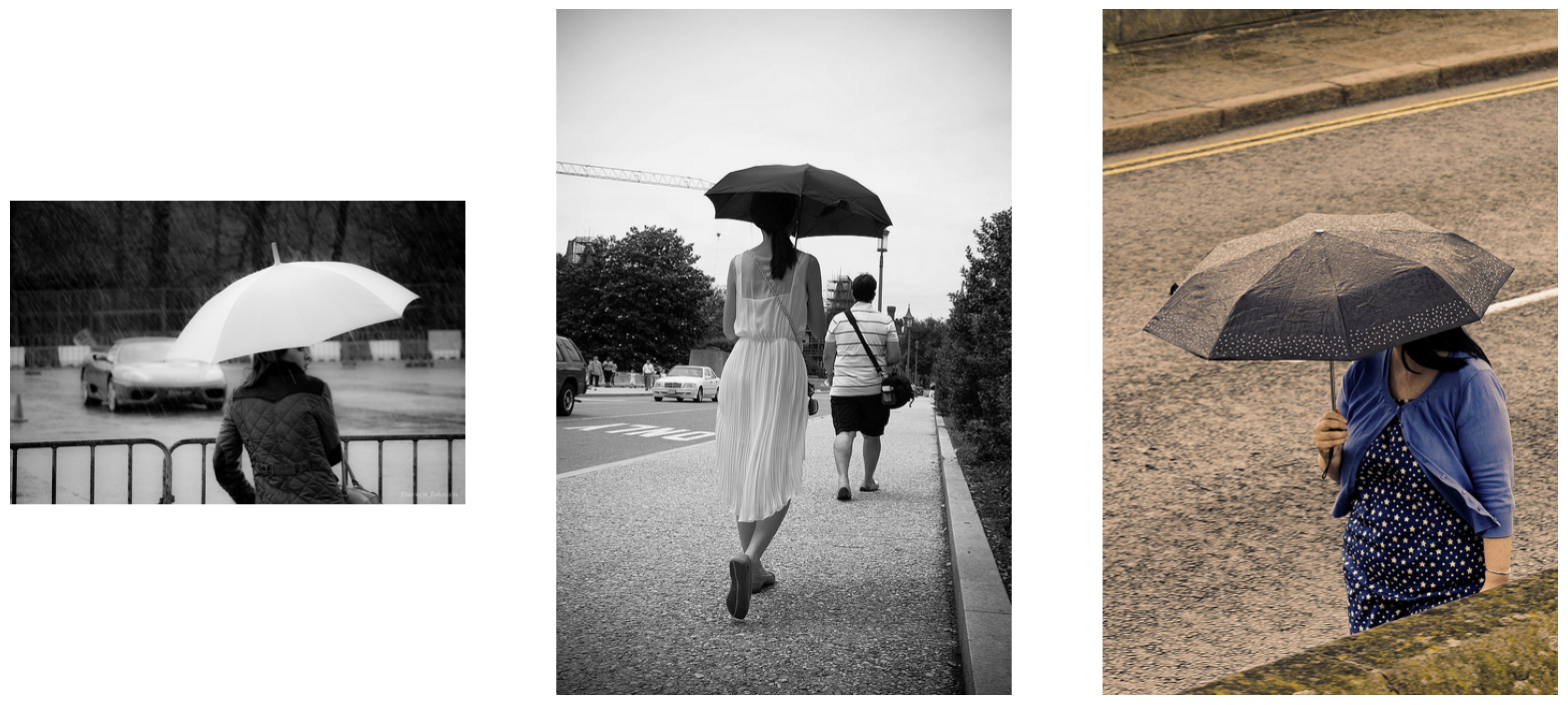}
\caption{Top-3 matching results for just semantic query}
\label{fig:complexsemantic}
\end{figure}

We compared the above SSQL query with just semantic query \textit{women with umbrella and two cars}. The results of this query are shown in the figure \ref{fig:complexsemantic}.

 \section{Discussion}
The proposed technique provides a new way to do semantic and structured searches at the same time on images. The current work can be extended in the following ways: 
\paragraph{Complex Queries} The current system supports only one semantic predicate. It can be extended to allow users to write complex queries and use the SEMANTIC predicate multiple times in a single query. For e.g.

\begin{lstlisting}
SELECT DISTINCT frame
FROM object_detection_results
WHERE (class = 'car' AND |\color{red}SEMANTIC| = "red color car") OR 
(class = 'bus'  AND |\color{red}SEMANTIC| = 'blue color bus');
\end{lstlisting}

\paragraph{Query Optimization} We assume that ML model results are already in the database. We preprocess the data to set up a vector store as well. It can be optimized to run ML models on demand depending on the query and caching the results in the database. Further, the query optimizer can optimize the order in which semantic query and SQL query are processed based on the cardinality estimates and cache hit rate.

\paragraph{Text datasets} We evaluated our system only on image datasets. It can be evaluated on text datasets as well without much modifications. 
 
\bibliographystyle{plain}
\bibliography{main.bib}

\end{document}